# Electric-field-tunable mechanical properties of relaxor ferroelectric single crystal measured by nanoindentation

Hao Zhou,[1] Yongmao Pei,[1,2,a)] Faxin Li,[1] Haosu Luo,[3] and Daining Fang[1,b)]

[1] *State Key Laboratory for Turbulence and Complex Systems, College of Engineering, Peking University, Beijing 100871, China*
[2] *State Key Laboratory for Strength and Vibration of Mechanical Structures, Xi'an Jiaotong University, Xi'an 710049, China*
[3] *Key Laboratory of Inorganic Functional Material and Device, Shanghai Institute of Ceramics, Chinese Academy of Sciences, Shanghai 201800, China*

Electric field dependent mechanical properties of relaxor ferroelectric material $Pb(Mn_{1/3}Nb_{2/3})O_3$-$PbTiO_3$ are investigated with the nanoindentation technique. Giant electric-field-tunable apparent elastic modulus (up to -39%), hardness (-9% to 20%) and energy dissipation (up to -13%) are reported. Based on experimental data, a characterization method of electromechanical coupled nanoindentation is proposed. In this method, an electric field tunable scaling relationship among elastic modulus, hardness and indentation work for ferroelectric materials can be determined. In addition, this method can be used to obtain the electric-field-dependent elastic modulus and hardness, and avoid the estimate of contact area in the Oliver-Pharr method. Finally, the different effects on elastic modulus between positive and negative electric fields can be explained by the flexoelectric effect.

a, b) Authors to whom correspondence should be addressed. Electronic mails: peiym@pku.edu.cn (Y. M. Pei), fangdn@pku.edu.cn (D. N. Fang).

Ferroelectric materials can be employed in functional devices such as ferroelectric memories,[1] piezoelectric actuators,[2] microwave devices[3] and energy harvesters.[4] The mechanical stress can significantly affect the performance and reliability of these devices. For example, most failures of multilayer actuators result from the stress concentration around the tips of internal electrodes.[5,6] On the macro-scale, the electric field dependent deformation behavior of ferroelectrics has been studied with the uniaxial compression method.[6-10] Schäufele and Härdtl found that the critical stress to cause depolarization of lead zirconate titanate (PZT) ceramics depended on the electric field.[6] Chaplya and Carman investigated the electric field dependent damping values and secant modulus of PZT ceramics.[7] McLaughlin et al. and Amin et al. reported the electric field dependent stress-strain curves of $Pb(Mn_{1/3}Nb_{2/3})O_3$-$PbTiO_3$ (PMN-PT) single crystals.[8,9] Li et al. found the large and electric-field-tunable superelasticity and damping factor in $BaTiO_3$ single crystals.[10] However, with the microminiaturization of ferroelectrics, traditional tension/compression test cannot satisfy the needs of mechanical testing at small scale. In recent years, nanoindentation technique is widely used due to its easy operation, high precision and low request of the sample preparation.[11-14] This technique avoids measuring the impression area under the microscope in traditional microhardness testing, but records the load and displacement signal during the entire indentation process and determines the mechanical properties through the mathematical analysis of the loading and unloading curves. Koval et al. studied the ferroelastic and piezoelectric behavior of PZT thin films with nanoindentation in the absence of external electric field.[15] Hysitron Inc. and Ruffell et al. developed the NanoECR module based on traditional nanoindentation technique, which could measure the indentation dependent contact resistance of ferroelectric films.[16,17]

In this work, with the multi-field nanoindentation technique[18] developed by us, the electric field dependent mechanical properties of ferroelectrics are investigated. The results show the giant tunable range of elastic modulus, hardness and energy dissipation. Based on experimental data, a characterization method of electromechanical coupled nanoindentation is proposed, which can be used to determine the electric-field-dependent elastic modulus and hardness, and avoid the estimate of contact area in the Oliver-Pharr method.

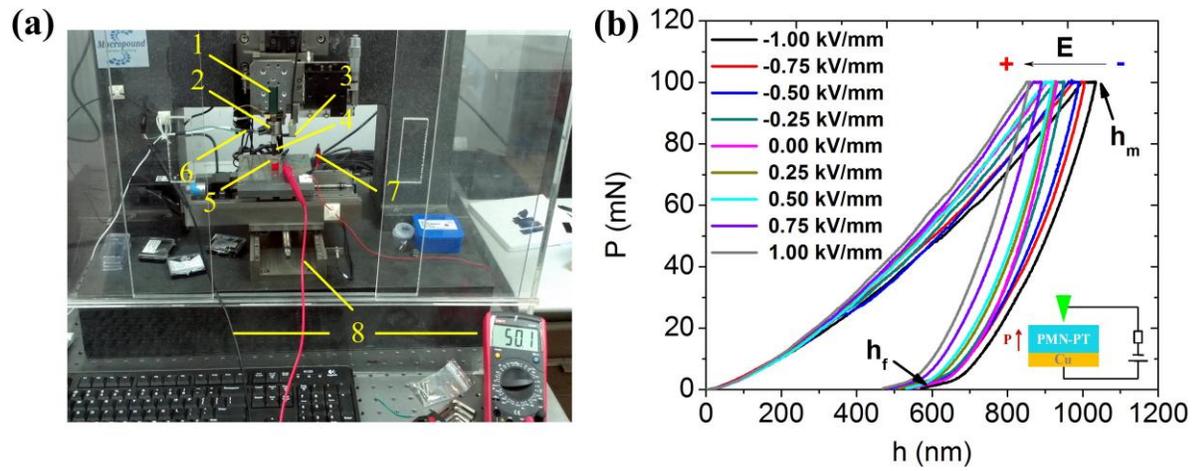

**Figure 1.** (a) Image of the experimental setup. 1. Piezoelectric actuator. 2. Load sensor. 3. Capacitive displacement sensor. 4. Conductive indenter tip. 5. PMN-PT on sample stage. 6. Negative/positive wire connecting the DC regulated power supply and the indenter tip. 7. Positive/negative wire connecting the DC regulated power supply and the bottom electrode of the sample. 8. Digital voltmeter monitoring the voltages between the indenter tip and the bottom electrode of the sample. (b) Nanoindentation load-depth curves with various electric fields.

The electromechanical coupled nanoindenter is composed of the mechanical module (piezoelectric actuator, load sensor, capacitive displacement sensor, and conductive indenter tip) and the electrical module (DC regulated power supply, positive and negative wires, conductive indenter tip, and bottom electrode of the sample), as Fig. 1(a) shows. The nanoindenter is carefully calibrated with standard sample (fused quartz), and the electric-field-independent mechanical properties of fused quartz manifests that the electric field has no influence on the performance of the instrument.[18] The tetragonal 0.62PMN-0.38PT crystal is poled and cut along the thickness [001] direction, with the dimensions of $5\times5\times0.2mm^3$ (l×w×t). The initial polarization direction is upward, as illustrated in the inset of Fig. 1(b). During the nanoindentation process, various DC voltages are applied to the sample via the upside conductive diamond tip (Berkovich, from SYNTON-MDP) and the downside copper electrode. The voltage ranges from -200V to 200V. The positive or negative sign means the electric field pointing upward or downward. Although the electric field inside the sample is inhomogeneous due to the asymmetric electrodes, here we use the nominal electric field strength (equal to the voltage divided by the sample thickness) to distinguish each other, as the treatment in the literature.[19] We conduct six indentations for each electric field, which add up to 72 tests for nine different fields. The typical indentation curves are shown in Fig. 1(b). It can be seen that both the positive and the negative electric fields affect the indentation curves greatly. With the electric field increasing from negative to positive, the maximum indentation depth $h_m$ and the final residual depth $h_f$ decrease. The curvatures of the loading and unloading curves are also influenced. It indicates that both the "elastic" and the "plastic" behavior of PMN-PT vary with the applied electric fields at the small-scale. Here, the "elastic" and "plastic" means the recovery and remnant deformation, accompanied by domain switching process, somewhat similar with the "elasticity" or "plastisity" of metals.[20]

In order to find the influence rules of electric fields on the mechanical properties of PMN-PT, several physical quantities are determined. The hardness is defined as $H_{IT} = P_{max}/A$, where $A$ is the projected area of the elastic contact, and $P_{max}$ is the maximum indentation load. The stiffness, $S = \mathrm{d}P/\mathrm{d}h$, is experimentally measured from the upper portion of the unloading curve. The effective elastic modulus, $M_r = \sqrt{\pi}S/(2\sqrt{A})$, is determined with the Oliver-Pharr method.[12] The energy dissipation is defined as the loading work minus the unloading work and then divided by the loading work, $\eta = (W_{load} - W_{unload})/W_{load} = \left(\int_0^{h_m} P\mathrm{d}h - \int_{h_f}^{h_m} P\mathrm{d}h\right)/\int_0^{h_m} P\mathrm{d}h$, where $W_{load}$ or $W_{unload}$ is the loading work or unloading work, which means the integral of the work done by the pressure to indent from 0 to $h_m$ or from $h_f$ to $h_m$. $h_m$ or $h_f$ is the maximum depth or final depth, as marked in Fig. 1(b). The results of the above physical quantities are shown in Fig. 2.

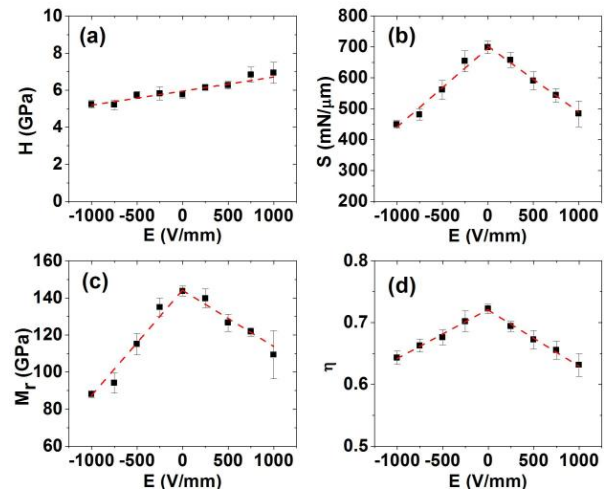

**Figure 2.** Electric field dependent (a) hardness; (b) unloading stiffness; (c) effective elastic modulus; (d) energy dissipation.

It can be seen from Fig. 2(a) that the indentation hardness shows a monotonic increasing relation (from -9% to 20%) with the electric field strength. Fig. 2(b,c,d) show that both the positive and the negative electric fields can reduce the unloading stiffness (up to -36%), the apparent effective elastic modulus (up to -39%) and the energy dissipation (up to -13%). The reduction becomes larger as the amplitude of the electric field, either positive or negative, increases. The fitting of experimental data gives a linear or piecewise linear empirical formula for each physical quantity, as follows:

$$H = cE + H_0, \tag{1}$$

where $H_0 = 5.76$ GPa, $c = 0.875$ kPa·m/V.

$$S = \begin{cases} b_1 E + S_0, & E < 0 \\ b_2 E + S_0, & E > 0 \end{cases}, \tag{2}$$

where $S_0 = 699$ mN/$\mu$m, $b_1 = 0.263$ N/V, $b_2 = -0.211$ N/V.

$$M_r = \begin{cases} a_1 E + M_{r0}, & E < 0 \\ a_2 E + M_{r0}, & E > 0 \end{cases}, \tag{3}$$

where $M_{r0} = 144$ GPa, $a_1 = 58.3$ MPa·m/V, $a_2 = -32.2$ MPa·m/V.

$$\eta = \begin{cases} d_1 E + \eta_0, & E < 0 \\ d_2 E + \eta_0, & E > 0 \end{cases}, \tag{4}$$

where $\eta_0 = 0.722$, $d_1 = 8.13 \times 10^{-5}$ mm/V, $d_2 = -9.24 \times 10^{-5}$ mm/V.

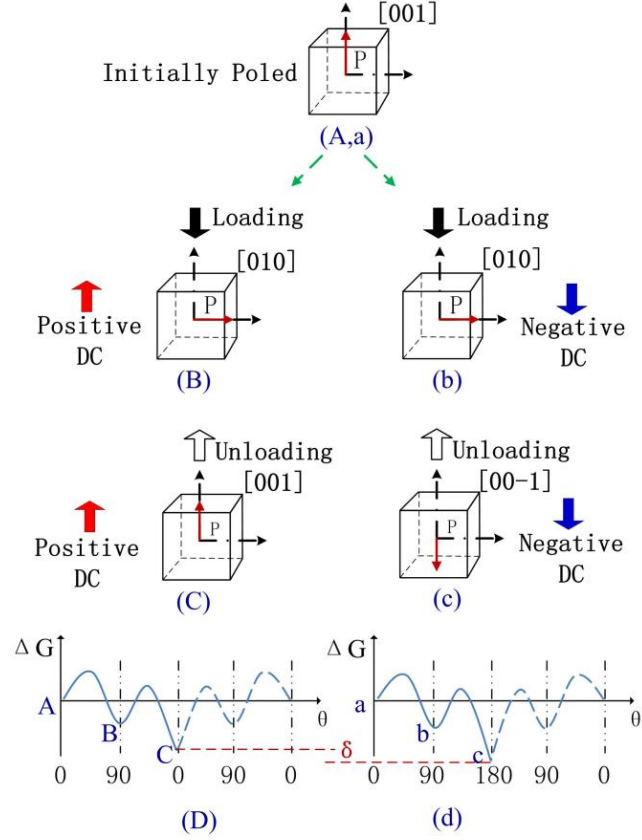

**Figure 3.** Schematic diagram of the domain switching process in the combination action of mechanical indentation and positive or negative electric field; and the corresponding potential well structures.

The electric tunable mechanical behaviors result from the competition between mechanical load and electrical field in influencing the domain switching process. During the loading upon the poled samples, the large compression stress in the thickness direction can induce the mechanical depolarization, i.e. switch the initially upward polarization to the in-plane direction via 90° domain switching as illustrated in Fig. 3(A-B, a-b), which contributes to the remnant deformation. Yet, the positive electric field can stabilize the initial polarization for the minimum of the electrostatic energy $G_e = -\int EP\mathrm{d}V$. In this case, it demands higher stress to cause the 90° domain switching. As a result, the average contact stress, i.e. the indentation hardness increases. On the contrary, the negative electric field destabilizes the upwards polarization, because $G_e = -\int EP dV > 0$ in this case, which makes it more prone to 90° domain switching. A lower stress can induce the mechanical depolarization process. Therefore, the average contact stress or the indentation hardness decreases. The linear relation between the hardness and the electric field of PMN-PT at the small-scale identifies qualitatively with the linear relation between the uniaxial compression coercive stress and the electric field of PZT at the macro-scale.[6] As mechanical loss accompanies domain switches, there is more irreversible work for negative voltages and less for positive voltages. Also, the loading work changes: there is more loading work for negative voltages and less for positive voltages. Based on the irreversible work and the loading work, the energy dissipation can be determined.

The electric field induced reduction of stiffness and apparent modulus can also be explained by the domain processes. During the mechanical unloading, the stress field inside the sample decreases rapidly. Therefore, some of the domains switched by the mechanical compression can be once again switched to the direction of the applied electric field via the second 90° domain switching as illustrated in Fig. 3(B-C, b-c). The associated switching of spontaneous strain constitutes the additional strain besides the pure elastic strain in the unloading process, i.e. $\varepsilon_{re} = \varepsilon_{ela} + \varepsilon_{dom}$. The higher the electric field, the more volume fractions of the second 90° domain switching, the larger the additional strain, and then the smaller the apparent elastic modulus and the stiffness.

Fig. 2(d) shows that both the positive and the negative electric fields can reduce the energy dissipation. The energy dissipation shows a piecewise linear relationship, which is different from the monotonic relationship between indentation hardness or elastic modulus and electric field strength. The proposed mechanism of energy dissipation in ferroelectrics is the domain process,[7] and it depends on the amount of domain switches and the ratio between reversible switches and irreversible switches. Comparing Fig. 2(d) with Fig. 2(c), we can see that the relationship between energy dissipation and electric field strength is similar to that between elastic modulus and electric field strength. Through dimensionless analysis of these physical quantities shown in Fig. 2, an electric field tunable scaling relationship among hardness, elastic modulus and indentation work is found, as shown in Fig. 4.

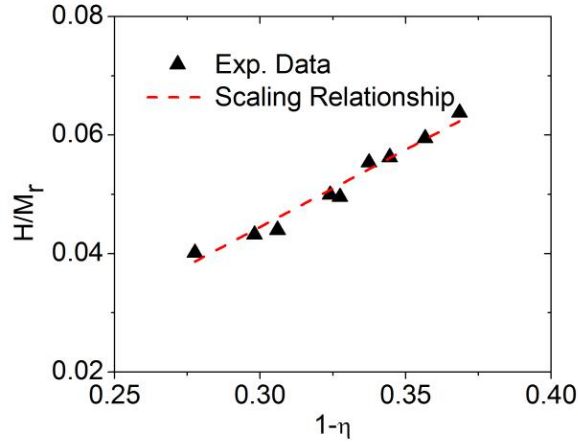

**Figure 4.** Electric field tunable scaling relationship among hardness, elastic modulus and indentation work with various electric fields. The experimental data are for various electric fields.

The black triangles in Fig. 4 are the experimental data for various electric fields. It shows that although the electric fields can change both the energy dissipation and the ratio of hardness to modulus, all these data distribute in one line, as the dash line in Fig. 4 shows, which is obtained by the linear fitting of the experimental data. This determines an electric field tunable scaling relationship:

$$H/M_r = k(1-\eta)+l = -dkE+(k+l-k\eta_0), \tag{5}$$

where $k = 0.2607$ and $l = -0.0337$ are the fitting parameters. The parameter $d=d_1$ or $d_2$ depends on the sign of electric field strength $E$ according to Eq. (4). Combining Eq. (5) and the definitions of hardness and elastic modulus, an energy method for determining the effective modulus and hardness is obtained:

$$M_r = \frac{\left[k(1-\eta)+l\right]\pi S^2}{4P_{max}} = \frac{\left[-dkE+(k+l-k\eta_0)\right]\pi S^2}{4P_{max}}, \quad (6)$$

$$H = \frac{\left[k(1-\eta)+l\right]^2 \pi S^2}{4P_{max}} = \frac{\left[-dkE+(k+l-k\eta_0)\right]^2 \pi S^2}{4P_{max}}. \quad (7)$$

This characterization method of nanoindentation for ferroelectric materials avoids the procedural error from estimating the contact area in the Oliver-Pharr method, and it is the development of the energy method for elastic-plastic materials.[21]

Finally, it must be noted that there are great difference in stress states between the small-scale indentation and the macro-scale uniaxial compression of materials. The former is in a high-stress state with an extremely large stress gradient due to the stress concentration, while the later is in a relatively low-stress state with the uniform stress field. The large stress gradient at the small scale can bring into focus another electromechanical coupling effect, i.e. flexoelectric effect (the coupling between strain gradient and polarization).[22-25] According to Sharma et al., the size effect of apparent elastic modulus measured by nanoindentation is most likely result from the flexoelectric effect.[22,23] Different from Sharma's approach by analyzing the stiffness per contact area at various indentation depths, we find in present work that the variation of apparent elastic modulus at constant indentation load with various applied electric fields can also indicate the possible existence of flexoelectricity. From Fig. 2c and Eq. (3), it can be seen that the slop of the fitting line for $E<0$ is larger than that for $E>0$, i.e. $|a_1|>|a_2|$. This means that the second 90 ° switching downwards by the negative electric field is easier than the switching upwards by the positive electric field. It can be explained from the perspective of energy qualitatively. The free energy of the ferroelectrics can be expressed as

$$\Delta G = \int \left[ G_p - \frac{1}{2}sX^2 - 2QXP^2 - \gamma(\nabla X)P - EP \right] dV \tag{8}$$

where $G_p = \frac{1}{2}\alpha_0(T-T_{0\infty})P^2 + \frac{1}{4}\beta P^4 + \frac{1}{6}rP^6$ is the polarization energy; $T$ is the temperature; $T_{0\infty}$ is the Curie–Weiss temperature of the bulk counterpart; $\alpha_0$, $\beta$ and $r$ are material parameters; $s$ is the elastic compliance coefficient; $Q$ is the electrostrictive coefficient; $X$ is the stress; $\nabla X$ is the stress gradient; $\gamma$ is the flexoelectric coupling coefficient. Among the five terms on the right hand of Eq. (8), only the fourth term is an odd function of polarization, $P$. The other four terms do not change sign if the applied field and polarization alter their signs concurrently. This means that the difference between the mechanical unloading response under the positive electric field (domain switching upwards) and that under the negative electric field (domain switching downwards) is caused by $\Delta G_{FxE} = \int (-\gamma(\nabla X)P) dV$, i.e. the flexoelectric coupling energy. For the sake of energy minimum ($\nabla X < 0$ and $\gamma > 0$ in this case), the polarization tends to point downwards. The stress gradient results in an asymmetrical energy curve with one well (corresponding to the negative polarization) deeper than the other (corresponding to the positive polarization), as shown in Fig. 3(D,d). The potential difference of state "C" and state "c" is $\delta = \Delta G_C - \Delta G_c = -2\gamma(\nabla X)|P| > 0$. That is to say, the stress gradient induced by the nanoindentation makes the downward polarization more stable, and therefore easy to occur, than the upward polarization. This conclusion are in accordance with the work of Lu et al., who found that the tip pressure destabilized the positive polarization by direct measurement of piezoresponse force microscopy (PFM) hysteresis loops.[24] In Lu's work, he also used the flexoelectric effect to rationalize their experimental results. In addition to flexoelectricity, an alternative contribution to the different effects on elastic modulus between positive and negative electric fields should be clarified. During loading, there are more 90° switches produced for negative electric field than for positive electric field, as a consequence of that the barrier for ferroelastic switching is lower when the external field opposes the internal polarization. Thus, during unloading, the starting situation is not the same for the positive electric field and for the negative one: the negative electric field unloads start from a situation where there are more horizontal domains to start with and therefore more domains that can be switched back to a vertical position. This would yield a bigger mechanical deformation response than for the positive electric field unloads, where there are less ferroelastic twins to start with and thus less potential ferroelastic switch-backs.

In summary, we have studied the electric field tunable deformation and energy dissipation behavior using the electromechanical coupled nanoindentation technique. The results show that the electrical tunable range of stiffness, apparent elastic modulus, hardness and the energy dissipation are quite large. Based on the nanoindentation results of ferroelectrics, a characterization method is proposed, which can be used to determine the electric field dependent elastic modulus and hardness, and avoid the estimate of contact area in the Oliver-Pharr method. In addition, an energy model taking into account the flexoelectric effect is used to explain the different effects on elastic modulus between positive and negative electric fields.

The authors are grateful for the support by the National Natural Science Foundation of China (11090330, 11090331 and 11072003) and the Chinese National Programs for Scientific

Instruments Research and Development (2012YQ03007502). Support by the National Basic Research Program of China (G2010CB832701) is also acknowledged.